\documentclass[aps,pre,twocolumn]{revtex4}
\usepackage{epsfig}

\begin{document}

\title{High-frequency flow reversal of AC electro-osmosis due to
  steric effects } 
\author{Brian D. Storey$^1$, Lee R. Edwards$^1$,
  Mustafa Sabri Kilic$^2$, and Martin Z. Bazant$^{2,3,4}$} 
\affiliation{
  $^1$ Franklin W. Olin College of Engineering, Needham, MA 02492  \\
  $^2$ Department of Mathematics, Massachusetts Institute
  of Technology, Cambridge, MA 02139 \\
  $^3$ Institute for Soldier Nanotechnologies, Massachusetts Institute
  of Technology, Cambridge, MA 02139\\
  $^4$ Physico-Chimie Th\'eorique, Gulliver-CNRS, ESPCI, 10 rue Vauquelin,
Paris  75005, France}

\date{\today}

\begin{abstract}
  The current theory of alternating-current electro-osmosis (ACEO) is
  unable to explain the experimentally observed flow reversal of
  planar ACEO pumps at high frequency (above the peak, typically
  $10-100$ kHz), low salt concentration ($1-1000$ $\mu$M), and
  moderate voltage ($2-6$ V), even if taking into account Faradaic
  surface reactions, nonlinear double-layer capacitance and bulk
  electrothermal flows. We attribute this failure to the breakdown of
  the classical Poisson-Boltzmann model of the diffuse double layer,
  which assumes a dilute solution of point-like ions. In spite of low
  bulk salt concentration, the large voltage induced across the double
  layer leads to crowding of the ions and a related decrease in
  surface capacitance. Using several mean-field models for
  finite-sized ions, we show that steric effects generally lead to
  high frequency flow reversal of ACEO pumps, similar to
  experiments. For quantitative agreement, however, an unrealistically
  large effective ion size (several nm) must be used, which we
  attribute to neglected correlation effects.
\end{abstract}

\maketitle

\section{Introduction}

Electrokinetic phenomena, which couple fluid flow, ion transport and
electric fields in electrolytes, are exploited in a variety of
microfluidic technologies~\cite{squires2005}.  A solid surface in
contact with an electrolyte typically acquires a surface charge and
forms an electric double layer composed of wall charge and a screening
layer of excess counterions.  Electro-osmosis results from the action
of an applied electric field on this double-layer charge, which sets
the bulk fluid into motion. Although the double layer is typically
very thin (nm) compared to the channel dimensions ($\mu$m), it is the
source of long-range flow patterns.

The classical continuum theory of
electro-osmosis~\cite{lyklema_book_vol2,hunter_book} suffices to
understand most direct-current (DC) electro-osmotic flows, which are
linear in the applied voltage. DC electroosmosis results from
application of an external electric field parallel to the surface,
which acts on equilibrium double-layer charge and induces motion in
the bulk, neutral liquid. The linear response is due to the assumption
that the applied field does not perturb the pre-existing surface
charge. When DC electroosmosis is used in microfluidic environments,
electrodes are placed at the ends of relatively long (cm) channels,
and therefore large voltages (kV) are typically needed to generate
sufficient fields.

\begin{figure}
\centering
  \epsfig{file=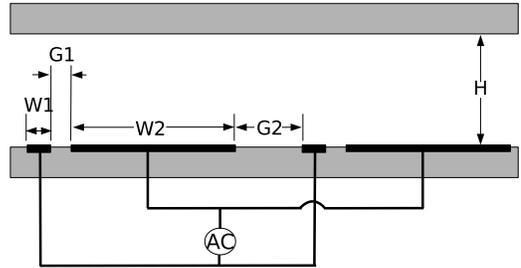,width=2.8in,clip=} 
  \caption{\label{fig:schematic} One period of an asymmetric planar
    array of microelectrodes in an AC electro-osmotic (ACEO)
    pump. This geometry has been studied experimentally by several
    groups~\cite{brown2001,studer2004,urbanski2006,microTAS2007} and
    is the subject of the present theoretical study.  The dimensions
    are; $W1=4.2~ \mathrm{\mu m}$ , $W2=25.7~\mathrm{\mu m}$, $G1=4.5~
    \mathrm{\mu m}$ , and $G2=15.6 ~\mathrm{\mu m}$, using the
    notation of Ref.~ \cite{olesen2006}.  }
\end{figure}

Electrokinetic phenomena can be exploited at much lower voltages and
with much greater precision by placing electrodes close together {\em
  inside} the fluid channels and using alternating current (AC) to
inhibit Faradaic reactions.  Large electric fields (100 V/cm) needed to
induce adequate flow can thus be generated with only a few volts by
micro-electrodes.  The most studied example is alternating-current
electro-osmosis (ACEO) \cite{encyclopedia_ACEO}, discovered by Ramos
et al. ~\cite{ramos1999,green2000b,gonzalez2000,green2002}.  When a
periodic array of interdigitated electrodes is placed inside the
channel, an AC signal applied to the electrodes can generate a steady
set of micro-vortices.  Ajdari showed that if geometrical asymmetry is
introduced, then these vortices can be rectified to drive net pumping
in one direction over the array~\cite{ajdari2000}. The original
implementation of this principle was a planar array of flat electrode
pairs of unequal widths and
gaps~\cite{brown2001,ramos2003,studer2004}, as shown in Figure
~\ref{fig:schematic}. Recent designs for ``3D ACEO'' pumps have
achieved much faster flows (mm/sec) using non-planar 
electrodes, which more efficiently rectify opposing slip velocities in
a ``fluid conveyor
belt''~\cite{bazant2006,urbanski2006,urbanski2007}. Waves of
voltage can also pump fluids over electrode arrays by traveling-wave
electro-osmosis (TWEO)~\cite{cahill2004,ramos2004}.

These effects exemplify the fundamental nonlinear electrokinetic
phenomenon~\cite{encyclopedia_nonlinear} of induced-charge
electro-osmosis (ICEO)~\cite{iceo2004a,iceo2004b,levitan2005}, which
also has other applications in microfluidics. The key source of
nonlinearity is that the diffuse double-layer charge is {\em induced}
near a polarizable surface by the applied electric field, which then
acts on it to drive nonlinear electro-osmotic flow. Although ICEO
flows were first described in the 1980s in the context of colloid
science~\cite{murtsovkin1996,gamayunov1986}, they are finding new
applications in microfluidic devices. As in the case of ACEO pumps,
broken spatial symmetries in ICEO flows can also be used to pump
fluids around polarizable microstructures and to manipulate polarizable
particles~\cite{iceo2004a,squires2006,yariv2006,velev}.

The current theory of ICEO flow~\cite{iceo2004b}, including
ACEO~\cite{gonzalez2000,olesen2006} and
TWEO~\cite{cahill2004,ramos2004} microfluidic pumps, is based on the
classical electrokinetic
equations~\cite{lyklema_book_vol2,hunter_book}. This century-old model
comprises the Poisson-Nernst-Planck (PNP) equations of ion transport
coupled to the Navier-Stokes equations of viscous fluid flow via
electrostatic Maxwell stresses. The crucial modification is in the
boundary conditions, relaxing the assumption of fixed surface charge
(or constant zeta potential) to allow for significant electrostatic
polarizability of the surface (e.g. fixed surface potential). The
electrokinetic equations themselves, however, have not been
questioned, until very recently~\cite{large}, which provides the
motivation for this work.

ICEO flows typically involve {\it large voltages} induced across the
double layer, greatly exceeding the thermal equilibrium voltage,
$kT/e$ ($=25$ mV at room temperature). In particular, experimental
ACEO electro-array
pumps~\cite{brown2001,studer2004,urbanski2006,urbanski2007,microTAS2007},
the subject of our study, usually apply several volts to the double
layer. Such large voltages of order $100 kT/e$ inevitably lead to the
breakdown of the PNP equations by violating the fundamental assumption
of an ideal dilute solution~\cite{kilic2007a,kilic2007b}.  The
limitations of dilute solution theory for double-layer structure are
well known, and many attempts have been made to incorporate effects
such as steric exclusion and electrostatic correlations.  (See
Ref.~\cite{kilic2007a} for a recent review.) ICEO flow, however,
raises new issues due to its more extreme, dynamical
context~\cite{large}. Even if the bulk solution is dilute, the surface
becomes so highly charged by the induced voltage that counterions
become crowded in the double layer, while simultaneously driving
tangential fluid flow.

\begin{figure}
\centering
\epsfig{file=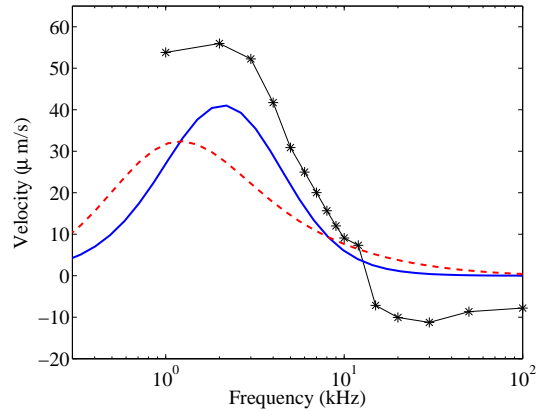,width=2.8in}
\caption{Pumping velocity vs. frequency for the planar ACEO pump in
  Figure~\ref{fig:schematic} from experiments and simulations.
  {\it Points:} Experimental data
  at 3 Vpp in 0.03 mM KCl from Ref.~\cite{microTAS2007} (taken by
  J. P. Urbanski); {\it Solid curve: } Linear Debye-H\"uckel-Stern
  model of the diffuse layer with a constant compact layer capacitance
  $C_s = \varepsilon/(\delta \lambda_D)$, where $\delta=0.5$ is chosen
  to fit the experimental peak frequency. {\it Dashed curve: }
  Nonlinear Gouy-Chapman-Stern (Poisson-Boltzmann) model of the
  diffuse layer also with $\delta=0.5$.  }
     \label{fig:jpdata}
\end{figure}

The need for improved models is also indicated by some unexplained
features of ICEO experiments. In this paper, we focus on the tendency
for planar ACEO pumps to reverse at high
voltage~\cite{studer2004,chang2004,wu2006,urbanski2006,microTAS2007},
which naturally poses problems for microfluidic applications, if it
cannot be reliably predicted. The standard model of ACEO based on the
classical electrokinetic equations predicts a single peak in the flow
rate versus frequency at the characteristic ``RC'' charging frequency
of the electrodes, in both the linear~\cite{ajdari2000} and
nonlinear~\cite{olesen2006,olesen_thesis} regimes.  Flow reversal has
been observed at high voltage ($> 2$ V) and high frequency (10-100
kHz) in ACEO pumping of dilute
KCl~\cite{studer2004,chang2004,microTAS2007} and deionized
water~\cite{urbanski2006} with $10 \mu$m scale electrode
arrays. 

Representative examples of the observed and predicted frequency
response are shown in Figure \ref{fig:jpdata} for the pump geometry of
Figure~\ref{fig:schematic}. Here we compare experimental data taken at
3.0 V $= 120 k T/e$ in an 0.03 mM KCl solution over gold electrodes
~\cite{microTAS2007} to simulations with two commonly used models: one
assumes the linear Debye-Huckel (DH) model of the double
layer~\cite{ramos1999,gonzalez2000,ajdari2000}, while the other adopts
a non-linear Poisson-Boltzmann (PB) model based on the Gouy-Chapman
solution~\cite{olesen2006}.  Following prior
work~\cite{ajdari2000,green2002,levitan2005}, a constant Stern-layer
capacitance is added in order to better {\em fit} the experimental data by
matching the peak frequency in the DH model.  
Both models can reasonably fit the data in terms of the 
peak in the forward pumping, though  
this becomes more difficult for PB at higher voltages \cite{olesen2006}. 
Each model predicts a single peak of forward flow,
while the experiments show the pump reversing above 10 kHz.  The
inability of these classical models to capture, even qualitatively,
this high frequency flow reversal motivates us to consider modified
double-layer models.

The reversal of ACEO pumps has been considered in several previous
studies. Flow reversal was first attributed to Faradaic
reactions~\cite{chang2004}, and this view of ``Faradaic charging''
persists in recent work on DC-biased AC
electro-osmosis~\cite{wu2007}. Simulations of ACEO pumps with
Butler-Volmer reaction kinetics, however, have failed to predict the
observed flow, especially at high
frequency~\cite{olesen2006,olesen_thesis}. In existing models, a weak
flow reversal due to reactions can only be observed at low frequency
(far below the maximum) and for certain sets of
parameters~\cite{ajdari2000,olesen2006,olesen_thesis}. This is
consistent with the observation of gas bubbles from electrolysis at
low frequency and high voltage in dilute KCl with gold
electrodes~\cite{studer2004}.  Recently, higher resolution
measurements of the same pump design with platinum electrodes have
revealed very weak ($<10 \mu$m/sec) reverse ACEO flow at low frequency
($< 20$ kHz) and low voltage ($<1.5$ V) and have demonstrated the
importance of Faradaic currents through {\it in situ} impedance
spectroscopy~\cite{gregersen2007}. Although the theory predicts
similar flow reversal, it is not in quantitative agreement and does
not predict the concentration dependence.
 
In addition to Faradaic reactions, various other nonlinear effects in
dilute solution theory also dominate at low frequencies: The
differential capacitance of the diffuse layer~\cite{lyklema_book_vol2}
diverges, which causes the RC charging time to grow exponentially with
voltage~\cite{olesen2006}, and salt adsorption and tangential
conduction by the diffuse layer are coupled to (much slower) bulk
diffusion~\cite{bazant2004,chu2006,chu2007}. All of these effects have
recently been incorporated in simulations of ACEO at large voltages
using the classical electrokinetic equations, but high-frequency flow
reversal was not observed~\cite{olesen2006,olesen_thesis}. Another
possible source of flow reversal is AC electrothermal
flow~\cite{gonzalez2006}, which can lead to reverse pumping in
experiments with planar electrode arrays~\cite{wu2007}, but under
different conditions of much higher salt concentration ($>$ 1 M),
voltage ($> 10$ V) and higher frequency ($> 100$ kHz) than ACEO ($<
0.01$ M, $< 5$ V, $< 100$ kHz).  To date, none of the above effects
have been able to explain the experimental data for ACEO pumps.

In this paper, we propose steric effects of ion crowding in the double
layer as a possible cause of the observed flow reversal at high
frequency.  Using simple modifications of the electrokinetic equations
to account for finite ion sizes, we are able to predict a frequency
response of ACEO pumps that is very similar to that observed in
experiments. Although our results do not provide a complete
quantitative theory, we will demonstrate that accounting for steric
effects can have an important impact on the theory of ACEO.

\section{Steric effects in a thin double layer}
\label{sec:steric} 

In the dilute-solution theory of
electrolytes~\cite{lyklema_book_vol2,hunter_book}, the concentration of
each ionic species in the diffuse part of the electric double layer at
a charged surface is in thermal equilibrium with a Boltzmann
distribution,
\begin{equation}
  c_i(x)=c_0 \mathrm{exp}{\left( \frac{-z_ie\phi(x)}{kT}\right)},  
\label{eq:cminus}
\end{equation}
where $k$ is Boltzmann's constant, $T$ the temperature, $e$ the
electron charge, $z_i$ the valence, $c_0$ the concentration in the
neutral bulk electrolyte just outside the double layer, and $\phi$ is
the (mean) electrostatic potential relative to the bulk ($x\gg
\lambda_D$, where $\lambda_D$ is the Debye-H\"uckel screening
length). In classical linear electrokinetic phenomena, the
diffuse-layer voltage drop $\phi(0)$ is set by chemical equilibrium at
the surface and is thus typically comparable to the thermal voltage
$kT/e$. At an electrode driving ACEO flow, however, much larger
diffuse-layer voltages $\phi(0) \gg kT/e$ are {\it induced} by a
applied voltage of order $100 kT/e$. Under these conditions, it is
easy to see that Boltzmann equilibrium (\ref{eq:cminus}) breaks down
by predicting diverging concentrations of counterions at the surface,
of order $e^{100}c_0$. 

The assumption of a dilute solution is thus incompatible with a large
applied voltage. If we assume ions have a characteristic length scale,
$a$, then the corresponding cutoff concentration $a^{-3}$ for the
breakdown of dilute-solution theory is reached at relatively low
voltage, even if the bulk salt solution is very dilute.  For example,
for $c_0=10^{-5}$ M, $z=1$, and $a=3$ \AA (including a solvation
shell), the cutoff concentration is reached at $0.33$ V.  To account
for the excess ions (at typical equilibrium voltages), Stern
postulated a compact layer of solvated ions of finite size on the
surface~\cite{bockris_book}, which carries most of the double-layer
voltage as the diffuse-layer capacitance diverges. Such an intrinsic
``surface capacitance'' is also invoked in models of ICEO flows, where
it can also include the effect of a thin dielectric coating on a metal
surface~\cite{ajdari2000,bazant2004}.  It seems unlikely, however,
that an atomically thin Stern or coating layer could withstand several
Volts, e.g. since dielectric breakdown occurs in most materials
(including water) in fields of order 10 MV/m $= 0.01$
V/nm~\cite{jones1995}. Instead, it seems the diffuse layer must carry
a substantial voltage $\phi \gg kT/e$ in ACEO experiments, which
causes the ions to become highly crowded near the surface.

\begin{figure}
(a)\includegraphics[width=2.8in]{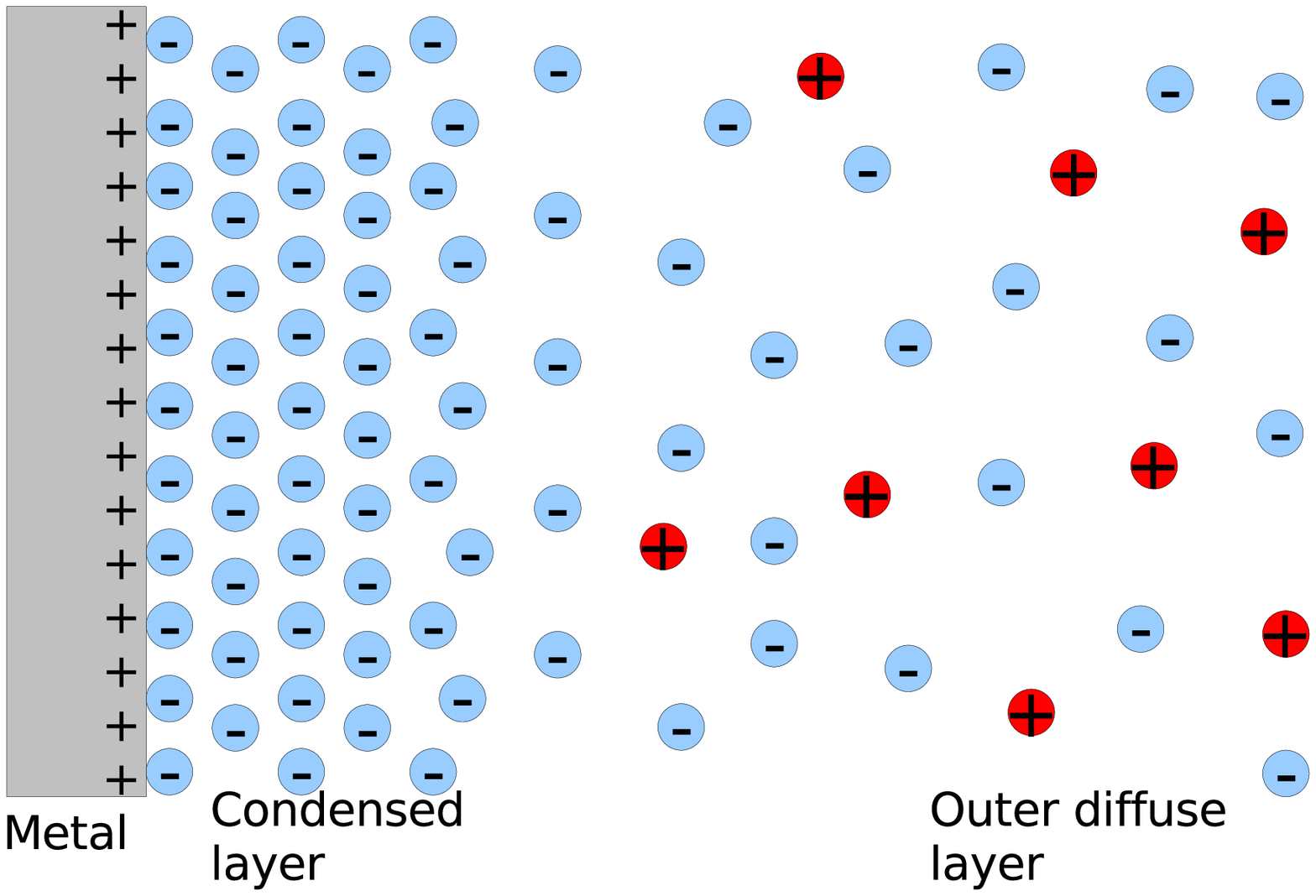}
(b)\includegraphics[width=2.8in]{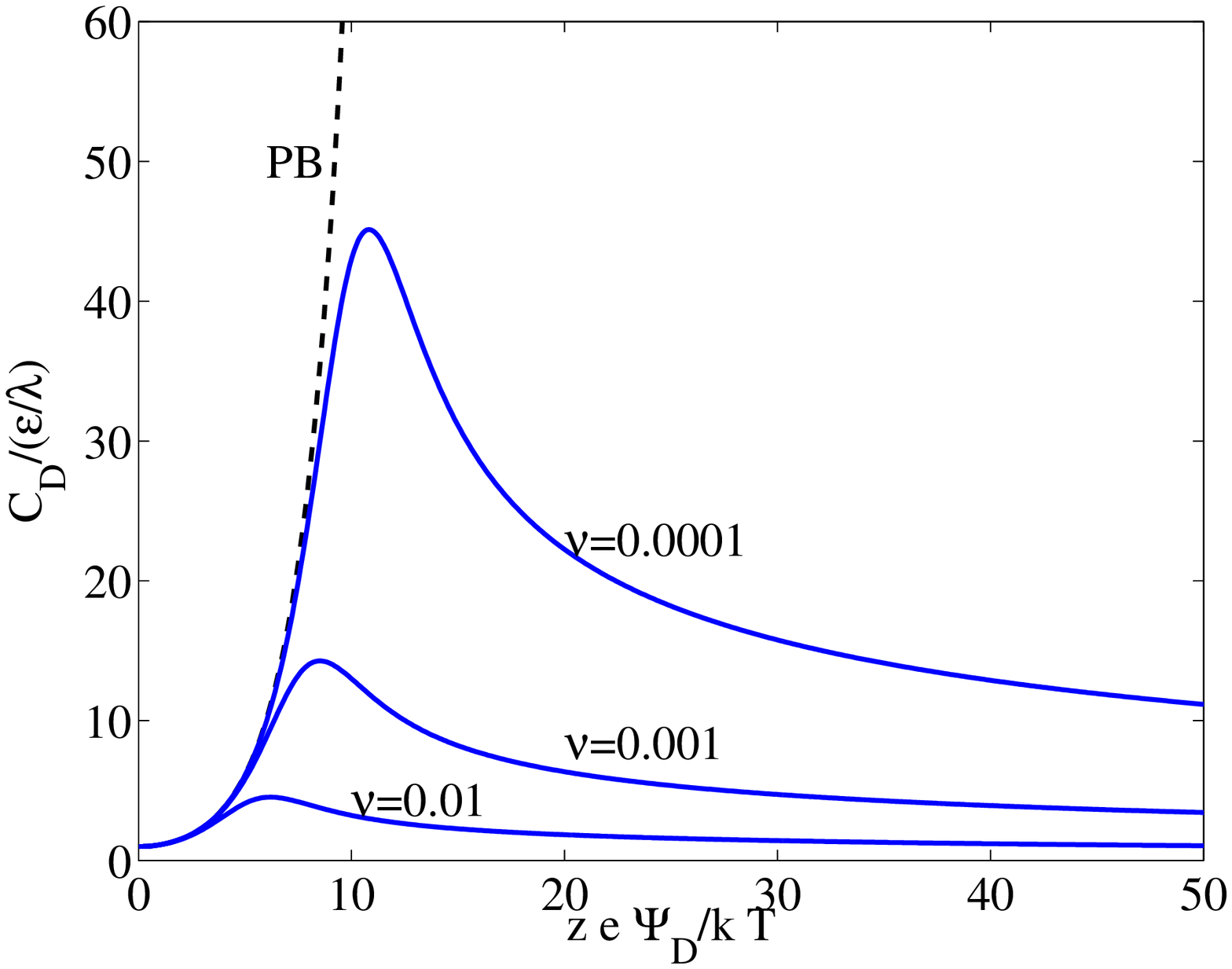}
\caption{\label{fig:C} (a) Schematic of the 
  equilibrium distribution of counterions
  near a positively charged surface 
  taking into account a minimum ion spacing $a$ for large applied
  voltages. As the voltage increases, the width of the condensed
  layer will increase causing a decrease the capacitance. 
  (b) The voltage dependence
  of the capacitance  of the diffuse 
  layer for PB and MPB at different values of $\nu$; 
  where $\nu = 2a^3 c_0$ is the
  bulk volume fraction of ions.  }
\end{figure}

A variety of ``modified Poisson-Boltzmann equations'' (MPB) have been
proposed to describe equilibrium ion profiles near a charged wall, as
reviewed in Ref.~\cite{kilic2007a}.  To capture ion crowding effects
across a wide range of voltages, we employ the simplest possible MPB
model~\cite{kilic2007a,kralj-iglic1996,borukhov1997,biesheuvel2007},
first proposed by Bikerman~\cite{bikerman1942},
which is a continuum approximation of the entropy of ions on a lattice
of size $a$. As shown in Fig.~\ref{fig:C}(a), when a large voltage is
applied, the counterion concentration exhibits a smooth transition
from an outer PB profile to a condensed layer at $c=c_{max}=a^{-3}$
near the surface.

In ACEO, charging dynamics are very important, so the double-layer
capacitance is an important property.  Dilute-solution theory predicts
that the differential capacitance, $C_D$, diverges with the voltage
as,
\begin{equation}
  C_{D}(\Psi_D)=\frac{\varepsilon_b}{\lambda_{D}}\cosh\left(\frac{ze\Psi_{D}}{2kT},
  \right) \label{eq:cdpb}
\end{equation}
where $\Psi_D$ is the voltage applied across the double layer.  For a
concentrated solution described by Bikerman's model, we predict the
opposite trend~\cite{kilic2007a},
\begin{equation}
C_{D}^{\nu }=
\frac{\frac{\varepsilon}{\lambda_{D}}\sinh(\frac{ze\Psi_{D}}{kT})}
{[1+2\nu\sinh^2\left(\frac{ze\Psi_{D}}{2kT}\right)]
\sqrt{\frac{2}{\nu} \mathrm{ln}[1+2\nu\sinh^2\left(\frac{ze\Psi_{D}}{2kT}\right)]}},
\label{eq:cdnu}%
\end{equation}
where $\nu=2 c_0 a^3$ is the bulk volume fraction of
ions~\cite{kilic2007a}. (Note that the same formula was derived by
Kornyshev~\cite{kornyshev2007} in the context of ionic liquids around
the same time that Kilic et al.~\cite{kilic2007a} derived it in the
present context of electrolytes.) 

As shown in Fig.~\ref{fig:C}(b), with this model the capacitance
reaches a maximum near the critical voltage then {\it decreases} at
larger voltages because the capacitor width grows due to steric
effects.  The same result also follows from a simpler model using PB
theory outside a condensed layer of variable width at the maximum
density ~\cite{kilic2007a}.  We will show that this general decrease
in capacitance with voltage that can drastically impact the charging
dynamics in ACEO flows.

As Biesheuvel {\it et al.} point
out~\cite{biesheuvel2005,biesheuvel2007}, Bikerman's latticed-based
approach can underestimate the true steric effect to a large degree,
and more accurate models of the entropy of a hard-sphere liquid, such
as the Carnahan-Starling (CS) equation of state or various extensions
to multicomponent mixtures~\cite{carnahan1969,lue1999}, can be used
instead.  The CS equation is not tractable for a closed-form
analytical expression for double-layer capacitance, but we can easily
obtain it numerically as a function of the voltage. For the details,
see the Appendix.

Although various mean-field models, such as CS, are very accurate for
equilibrium properties of the hard-sphere gas and liquid at low to
moderate volume fractions, $\Phi < 0.55$, none can claim to accurately
describe the packing limit at large volume fractions, which would
generally lead to the decrease in double-layer capacitance at large
voltage~\cite{kilic2007a}. For example, these models predict a
diverging chemical potential only at $\Phi=1$, and not and not at the
mathematical packing bound ($\Phi=0.74$, corresponding to the
face-centered cubic lattice) or the more relevant bound of the jamming
transition (around $\Phi=0.63$ for hard spheres
~\cite{torquato00}). Even in simple model systems in equilibrium, the
jamming transition is poorly understood
~\cite{liu98,liu02,ohern02,ohern03,donev04}, so none of these models
can possibly be expected to accurately capture the dynamics of jamming
of hydrated ions very close to a surface driven by a large AC voltage,
while also producing ACEO flow. Of course, under such extreme
conditions, the mean-field approximation for both steric and
electrostatic interactions should also be questioned.

Due to all of these complexities, following Bazant {\it et
  al.}~\cite{large}, we will focus on the simplest approximation of
Bikerman's model in order to explore the qualitative impact of steric
effects on ACEO flow. It will turn out that this very simple model
suffices to predict high frequency flow reversal of ACEO pumps. We
will then repeat some calculations with the CS model, which improves
the agreement with experiments somewhat, but does not alter the
qualitative predictions of Bikerman's model.

\section{Model for ACEO pumps}

In our theoretical study of ACEO pumping, we consider the standard
experimental geometry of planar, asymmetric electrode-array pumps
shown in Fig. \ref{fig:schematic} due to Brown {\it et
  al.}~\cite{brown2001}.  The specific dimensions are selected to
match the experiments Studer \textit{et. al} \cite{studer2004} and
Urbanski {\it et al.}  ~\cite{urbanski2006,microTAS2007}.  The basic
question of ACEO flow in such pumps has been thoroughly addressed by
Olesen {\em et. al} ~\cite{olesen2006,olesen_thesis} using the
classical electrokinetic equations of dilute solution
theory~\cite{lyklema_book_vol2,hunter_book}.  The significant
difference between this work and Ref. \cite{olesen2006} is our
application of the new physical model for the double-layer
capacitance, taking into account steric effects at large
voltage. Unlike Ref.~\cite{olesen_thesis}, we also ignore ``strongly
nonlinear'' effects associated with salt uptake by a highly charged
diffuse layer and resulting bulk diffusion~\cite{bazant2004,chu2006},
since these phenomena are greatly reduced by steric
effects~\cite{kilic2007a,kilic2007b}.

In the bulk fluid outside the
electric double layers, we assume electroneutrality with a constant
electrical conductivity, $\sigma$.  Under these assumptions, the
electric potential in the fluid bulk, $\phi$, satisfies Laplace's
equation, which is equivalent to Ohm's law for a constant bulk
resistance. This corresponds to the ``weakly nonlinear'' regime where
nonlinear circuit models still
hold~\cite{bazant2004,chu2006,olesen_thesis}.  For boundary conditions
on the potential, we assume that there is no normal Ohmic current on
the sections of channel substrate at $y=0$ and the upper boundary at
$y=H$.

For the boundary condition over the electrodes, we must incorporate a
model of the thin electric double layers that will form.  These double
layers are so thin (10-100 nm), that when we apply the boundary
condition on $\phi$ at $y=0$, we assume this is outside the double
layer and not at the electrode surface.  The local charge stored in
the double layer, $q(x)$, changes in time due to the Ohmic current.
Our double layer model provides the voltage difference between the
fluid bulk and the externally applied voltage at the electrode,
$V_{ext}$, based on the amount of charge stored.  Thus the boundary
condition for the electric potential in the bulk is dynamically
determined.  While Faradaic current certainly play a role in ACEO, we
neglect it in our model to focus on steric effects alone, which have
not previously been considered. The role of reactions in classical
double-layer models in ACEO was studied by Olesen {\em
  et. al}~\cite{olesen2006} and shown not to play a role in
high-frequency flow reversal.

As in prior work~\cite{ajdari2000,olesen2006}, we introduce the following
scales for the nondimensionalization of the governing equations,
\[
    \quad[x,y] = L,\quad[t] = \frac{L C_o}{\sigma}, \quad[\phi ]=\frac{k_B T}{z e},\quad [q]=\frac{k T }{z e }C_o.
\]
Note that $[t]$ is the ``RC time'' for the equivalent
circuit~\cite{ramos1999,bazant2004}, where the characteristic Debye
capacitance is $C_o = \frac{\varepsilon}{ \lambda_D}$ where
$\varepsilon$ is the permittivity of the electrolyte and $\lambda_D$
is the Debye length.  The characteristic length scale for computing
the bulk resistance, $L/\sigma$, is the electrode spacing $L=G_1$.  We
assume no Stern layer with our new double layer model, though its
inclusion would be trivial and would not significantly alter our
results.

The dimensionless problem reduces to Laplace's equation for the
potential in the fluid bulk,
\begin{equation}
\nabla^2 \phi = 0. 
\label{eq:lap}
\end{equation}
The boundary condition on the 
substrate (the entire upper wall at $y=H/L$ and the insulating 
regions along $y=0$ between the electrodes) is
\begin{equation}
\frac{\partial \phi}{\partial y}=0, 
\label{eq:bc1}
\end{equation}
The boundary condition over the electrode double layer
is given by, 
\begin{equation}
  \phi =    V_{ext} - \Psi_D(q), 
\label{eq:bc2}
\end{equation}
where  $\Psi_D(q)$ is the 
total potential drop across the  double layer. The functional 
relationship between   the double layer voltage drop and charge
is given by the 
model from Kilic \textit{et al.} \cite{kilic2007a},
\begin{equation}
\Psi_D =  - 2~ \mbox{sign}(q)~    \mbox{sinh}^{-1}  \left( \sqrt{ \frac{1}{2 \nu} \left(  \mbox{e}^{\frac{q^2 \nu}{2}}-1 \right)}  \right),
\label{eq:q_full}
\end{equation}
while the local charge is dynamically determined by the ``RC''
condition,
\begin{equation}
\frac{dq}{dt} = \left. \frac{\partial \phi}{\partial y} \right|_{y=0}.
\label{eq:charge}
\end{equation}

The closed set of equations (\ref{eq:lap})-(\ref{eq:charge}) form our
electrical model.  Note that in the limit of very small $\nu$ in
Eq. \ref{eq:q_full} we obtain the classic result from
Poisson-Boltzmann, $\Psi_D = -2~ \mbox{sinh}\left(\frac{q}{2}
\right).$ In the limit of small potential and small $\nu$ we obtain
$\Psi_D = -q,$ the linear Debye-Huckel model.  The later two limits
have been extensively studied \cite{olesen2006}; it is the impact of
Eq \ref{eq:q_full} on the flow that is the focus of this paper.

Once the electrical problem is solved, we compute the electroosmotic
slip by the Helmholtz-Smoluchowski relationship,
\begin{equation}
U =   \Psi_D \frac{\partial \phi(y=0)}{\partial x}.
\label{eq:velocity}
\end{equation}
and ignore any modifications at large voltage, e.g. due to crowding
effects~\cite{large}, in order to focus on steric effects in the
double-layer charging dynamics. Velocity is made dimensionless by the
natural scale~\cite{ramos1999,iceo2004a},
\[
[U] = \frac{\varepsilon V^2}{\eta L}
\]
where $V$ is the magnitude of the applied voltage and $\eta$ is the
viscosity.  An interesting fact, exploited by previous
authors~\cite{ajdari2000,gonzalez2000,olesen2006}, is that we can
compute the net pumping without actually solving the flow field in the
channel, but by taking the time and spatial average of equation
\ref{eq:velocity} along the boundary $y=0$.  Further details of this
derivation can be found elsewhere, e.g. in Olesen {\em
  et. al}~\cite{olesen2006}.

Due to the irregular electrode geometry and the non-linear charging
processes, we must resort to solving Equations
(\ref{eq:lap})-(\ref{eq:charge}) numerically.  We couple the solution
of Laplace's equation to the dynamic boundary condition for the
charge.  The solution is integrated forward in time until an adequate
steady state is reached and the net flow is computed by averaging
Eq. \ref{eq:velocity} in time and space. Details on the numerical
method can be found in the supplementary material \footnote{ See EPAPS
  Document No. XXXX for details of the numerical method. For more
  information on EPAPS, see http://www.aip.org/pubservs/epaps.html }.
The simulations were confirmed by an independent method using the 
commercial software package COMSOL Multiphysics.

\section{Results}

\begin{figure}
\centering
\epsfig{file=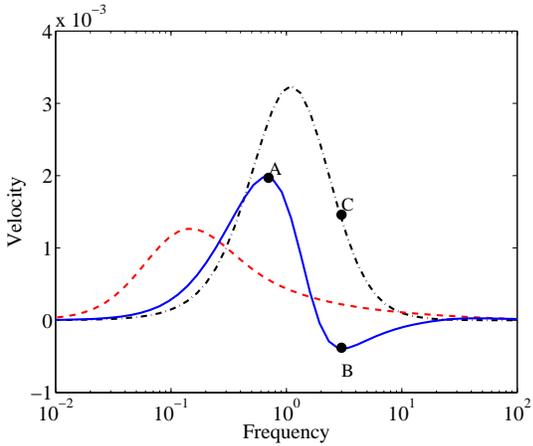,width=2.8in}
\caption{Average slip velocity as a function of frequency for three different
  models at 2.5 Volts.  The solid blue curve is for $\nu=0.01$ with
  the complete model, the black dash-dot line assumes the linear
  Debye-Huckel relationship $q=-\Psi_D$, the red dashed curve is for
  $\nu=0$ with a Stern layer capacitance of $\delta=0.1$, Only the
  model with steric effects predicts flow in the negative
  direction. Points A, B, and C will be referred to in
  Fig. \ref{fig:flow}.  }
     \label{fig:models}
\end{figure}

We start by comparing our results with the steric model for the double
layer, Eq. (\ref{eq:q_full}), to simulations of the dilute-solution
models used in all prior work. In Fig. \ref{fig:models}, we compare
the flow velocity as a function of frequency for the classic linear
Debye-Huckel model valid only when the applied voltage is much less
that $k T/e$, the Poisson-Boltzmann non-linear capacitance which is
recovered from our model when $\nu=0$, and the model which accounts
for steric effects, Eq. \ref{eq:q_full}.  At a voltage of $100
kT/e=2.5$ V, we see that the model which accounts for steric effect
shows a high frequency flow reversal.  As we note in
Ref.~\cite{large}, to our knowledge, these results represent the first
demonstration of a mathematical model that can predict such reversal
of flow in ACEO pumps.

\begin{figure*}
\centering
  \epsfig{file=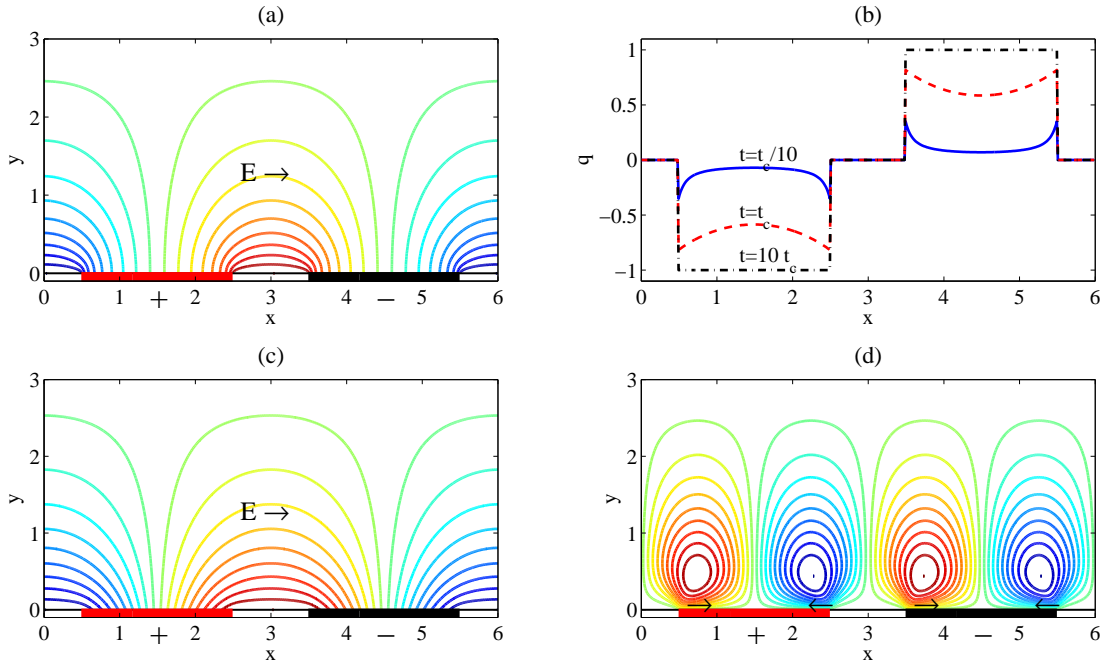,width=5.8in,clip=} 
  \caption{\label{fig:mechanism} Basic mechanism for flow in ACEO
    devices. Here we consider a symmetric array of electrodes under a
    suddenly applied voltage.  In (a) we show electric field lines at
    the instance the field is turned on. The electric field points
    from the positive to the negative electrode.  The location on the
    electrodes at $y=0$ is shown.  In (b) we show the charge density
    over the electrodes at $1/10$ the charging time $t_c$, at the
    charging time, and at $10$ times the charging time. In (c) we show
    streamlines of the electric field at the charging time. In (d) we
    show electric field lines at the charging time. The arrows denote
    the flow direction.  }
\end{figure*}

In order to understand the physical mechanism of flow reversal, we
must first understand why there is flow at all.  The basic mechanism
for flow in ACEO can be best understood when considering a
geometrically symmetric electrode array, as in Fig
\ref{fig:mechanism}, subjected to a suddenly applied DC voltage
(without any Faradaic current) ~\cite{encyclopedia_ACEO}.  When the
voltage is initially switched on, the electric field lines point from
the positive to the negative electrodes as shown in Fig
\ref{fig:mechanism}(a).  Since there is initially no double layer
charge, the field lines are perpendicular at the electrode surface.  A
current will begin to flow through the electrolyte with ions migrating
along the electric field lines and the double layers over the
electrode will begin to charge as shown in Fig \ref{fig:mechanism}(b).
The electric field is strongest at the electrode edges and the
resulting high current quickly charges the double layer at that
location.  On the order of the characteristic charging time, $t_c$,
(as computed using the gap length scale) the electrode edges will be
significantly screened.  In Fig \ref{fig:mechanism}(b) we show the
charge at $y=0$ at three instances in time.

As the double layers charge up, the electric field lines in the fluid
bulk are diverted around the screened electrode edges, resulting in a
tangential component of the electric field.  This is shown in Fig.
\ref{fig:mechanism}(c) where we can see the field lines are no longer
perpendicular over the entire electrode, especially near the electrode
edges.  The tangential electric field exerts a force on the fluid in
the charged double layer, causing flow.  Given the direction of the
field and sign of the charge, the flow always goes from the edge of
the electrodes, inward.  The flow streamlines and flow direction are
shown in Fig. \ref{fig:mechanism} (d).  If the applied voltage is
maintained, eventually the entire electrode becomes fully screened
there is no electric field in the bulk and the flow will stop.  The
above argument about the direction of flow did not depend upon the
sign of the applied voltage, reversing the sign causes the same flow
pattern.  Therefore, if an AC voltage is applied at a frequency
corresponding to the charging time, time averaged vortex flow can be
maintained.

\begin{figure*}
\centering
  \epsfig{file=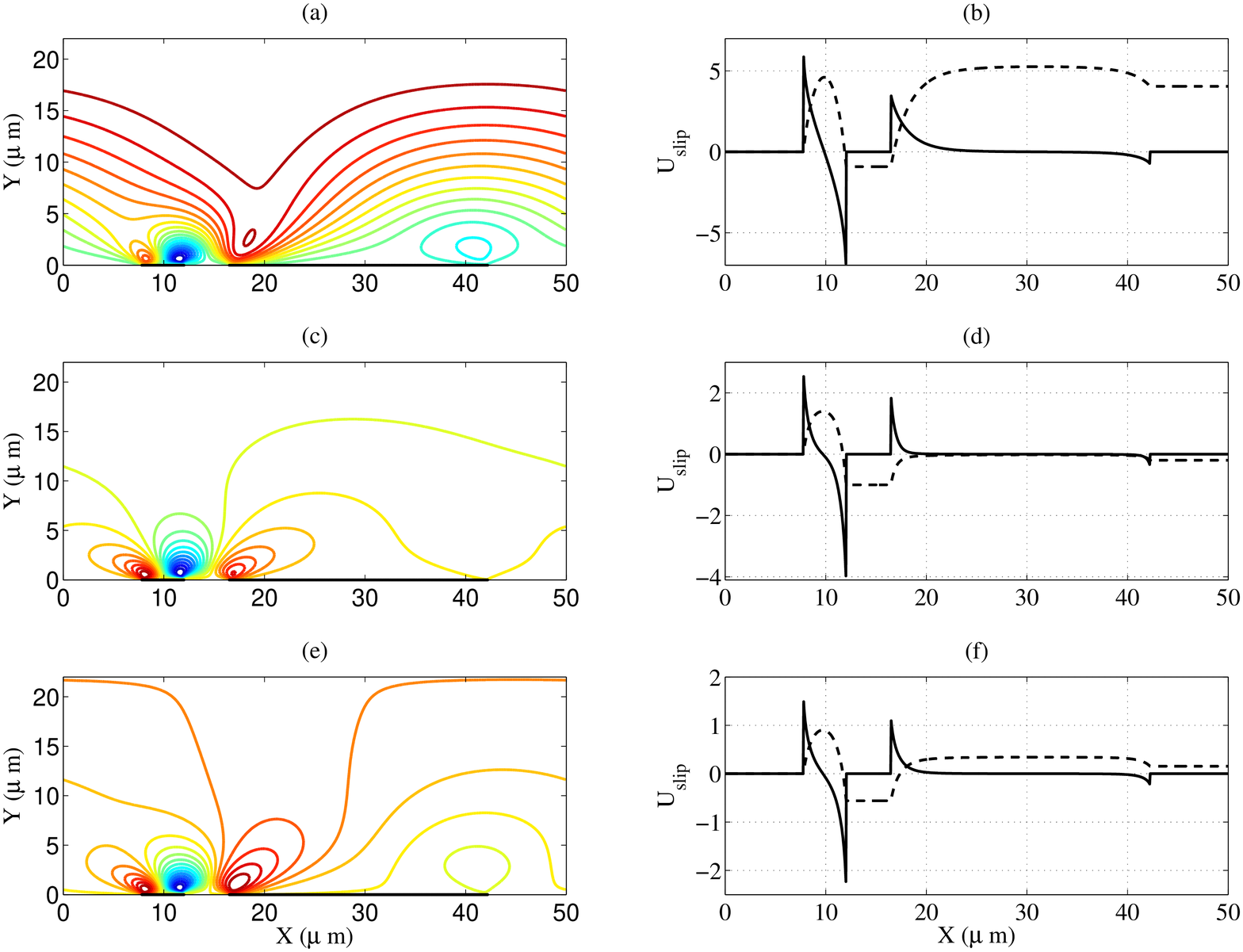,width=5.8in,clip=} 
  \caption{\label{fig:flow} Streamlines (a,c,e) and slip velocity
    (b,d,f) for three different cases. The first row is for $V=100$,
    $\nu=0.01$, and $\omega = 0.7$. This frequency corresponds to the
    peak in the forward flow for the steric model. In (a) we show the
    streamlines. In (b) we show the time-averaged slip velocity (solid
    line) and the cumulative spatial integral of this velocity (the
    dashed curve). The second row, figures (c) and (d) is for $V=100$,
    $\nu=0.01$, and $\omega = 3$. This frequency corresponds to the
    peak in the reversed flow. The third row, figures (e) and (f) is
    for $V=100$ and $\omega = 3$ with the linear Debye-Huckel model
    $\Psi_D=-q$.  }
\end{figure*}

Ajdari introduced the idea of rectifying this symmetric flow pattern
to create an ACEO pump by introducing various forms of geometrical
asymmetry within each spatial period of the electrode array
\cite{ajdari2000}. This principle was first implemented by Brown {\it
  et al.} using flat, co-planar electrodes of different widths and
gaps~\cite{brown2001}, as shown in Fig. \ref{fig:schematic}, and this
general design has been studied extensively by several
groups~\cite{ramos2003,mpholo2003,studer2004,olesen2006,
olesen_thesis,gregersen2007,urbanski2006,microTAS2007}.
In this geometry, the symmetry breaking causes the flow to go from
left to right by upsetting the delicate balance between leftward and
rightward slip patterns. The fact that pumping comes from a relatively
small bias of opposing, co-planar surface flows makes this design
susceptible to flow reversal if nonlinearities in charging dynamics
tip the balance the other way, as we demonstrate below. Current
designs using stepped, three-dimensional electrodes are much faster
and more robust against flow reversal, since the opposing slip
patterns work together to drive pumping in one direction, as a ``fluid
conveyor
belt''~\cite{bazant2006,urbanski2006,urbanski2007,burch_preprint}. The
sensitivity to flow reversal in the planar pump, however, allows us a
better opportunity to explore subtle nonlinearites in double-layer
charging dynamics, which is why we have chosen it for our theoretical
study.

In Fig. \ref{fig:flow} we demonstrate how the geometrical asymmetry
results in forward flow when driven at the charging time and how
steric effects can produce high frequency reversal. In
Fig. \ref{fig:flow}(a) and (b) we show the time averaged streamlines
(a) and time average slip velocity (b) for the ACEO pump driven at the
charging time, corresponding to point A in Fig. \ref{fig:models}.  We
notice a few features of the slip profiles.  First, the magnitude of
velocity over the small electrode is greater than the velocity over
the large electrode. This difference is due to the fact that the {\em
  total} charge contained on both electrodes must be the same, the
local charge density (and thus electroosmotic slip) is always higher
on the small electrode.  Second, we notice slip profiles over either
electrode is asymmetric with the greatest slip found at the edges at
the small gap. This effect is simply due to high electric field in the
small gap which can exert a larger force.  Finally we see the
asymmetry in the slip is greater for the large electrode than the
small electrode, an effect due to the geometry.  The result for this
geometry and frequency is the net flow goes in the positive $x$
direction.  The relative contributions to the flow is easily seen by
the cumulative spatial integral of the time averaged slip; the dotted
line in Fig. \ref{fig:flow} (b).

In Fig. \ref{fig:flow} (c) and (d) we show the behavior at the
approximate peak of the reverse flow in Fig. \ref{fig:models},
corresponding to point B in Fig. \ref{fig:models}.  The same generic
features of the slip profiles are the same as in (b). However, the
relative contributions have changed at this frequency.  The reason for
the reversal can be explained by the change in charging time for the
different sized electrodes.  When steric effects are included, we must
remember that the capacitance decreases as more charge is stored
inside the double layer.  Since the charge density over the small
electrode is much greater, the capacitance of the small electrode's
double layer is {\em decreased} at high voltage due to the steric
effects. The lower capacitance means that the charging time of the
small electrode is decreased relative to the large electrode. This
change in charging time means that at high frequency and high voltage
the small electrode has more time to sufficiently charge to generate
flow.

This idea is illustrated in Fig. \ref{fig:flow} (c) and (d).  The
cumulative spatial integral of the time-averaged slip shows that the
negative flow contribution from the small electrode is nearly twice
the positive contribution. At this high frequency the large electrode
is not very highly charged and its forward flow component adds only a
small amount to the net.  That this reversal is caused by the change
in the charging dynamics can be seen in Fig. \ref{fig:flow} (e) and
(f), which shows the same condition as (c) and (d) only with the
linear Debye-Huckel model; point C in Fig. \ref{fig:models}.  Here,
since the large electrode has the same charging time as the small
electrode, its forward component is significant to provide a net
positive flow.  It is clear from these pictures that with the
inclusion of steric effects, the direction and magnitude of flow in
ACEO pumps is determined by a delicate balance influenced {\em both}
by the geometry and the charging dynamics.

\begin{figure*}
\centering
\epsfig{file=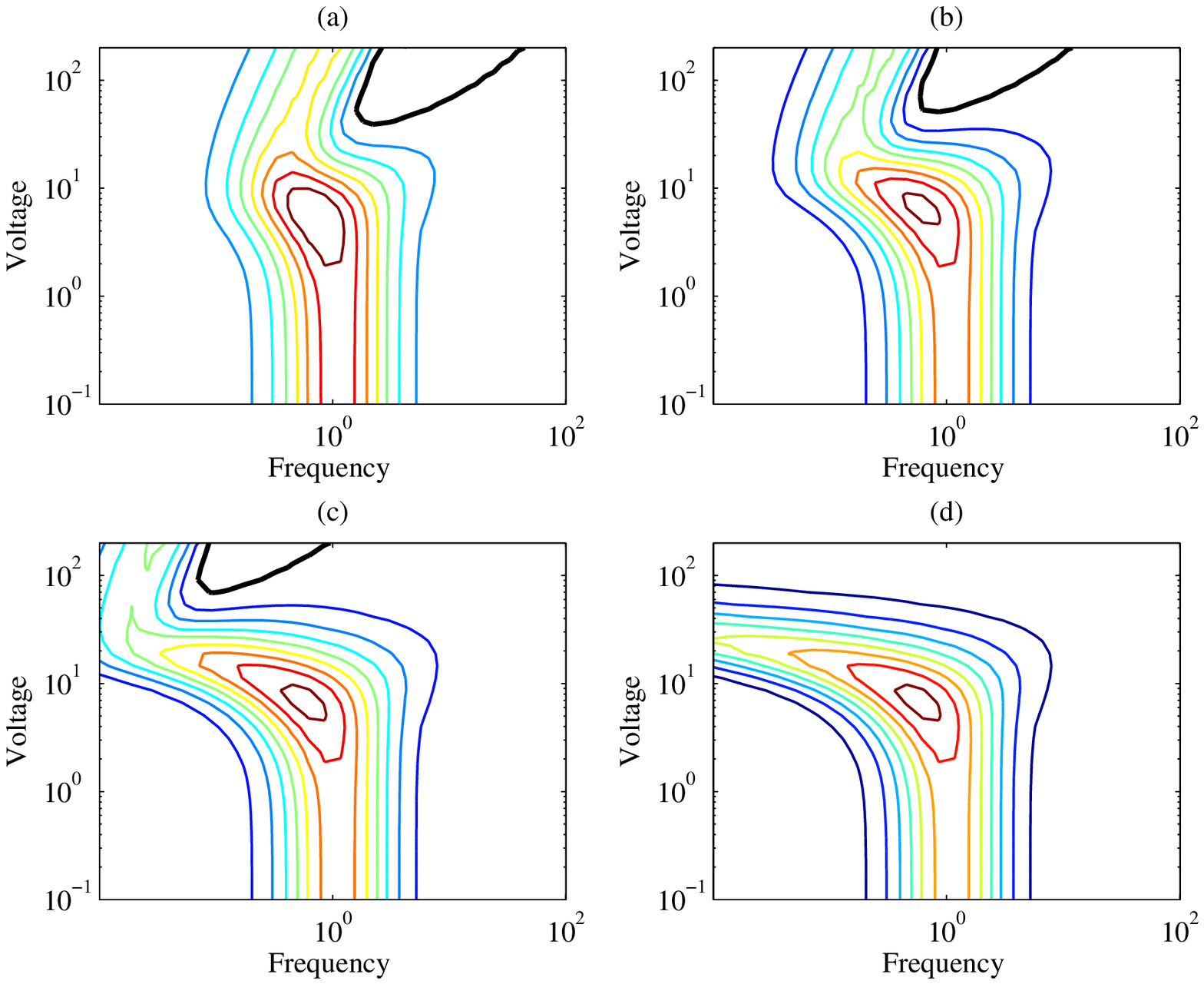,width=5.8in,clip=}
\caption{Flow response as a function of voltage for various values of
  $\nu$, the steric parameter. We show values of $\nu=0.01$ (a),
  $\nu=0.001$ (b), $\nu=0.0001$ (c), $\nu=0$ (d), The contours show
  constant levels of total flow rate. The heavy line shows the $U=0$
  contour and the region above this line denotes the region where flow
  reversal occurs. }
     \label{fig:steric_nu}
\end{figure*}

Up to now, we have taken the volume fraction, $\nu$, to be a constant.
To better see the effect of $\nu$ on the predicted flow, in
Fig. \ref{fig:steric_nu} we show contours of the flow rate as a
function of voltage and frequency at different values of $\nu$.  For a
fixed physical value of $a$, different values of $\nu$ represent how
the flow would change as a function of concentration (in dimensionless
terms).  The heavy lines denote the region where the flow is negative,
indicating a reversal of flow.  We see that the frequency response of
the flow does not change with voltage at low voltage, as would be
predicted by the linear Debye-Huckel model.  At low voltage, we see
only a single peak in the forward flow at the characteristic
frequency.  As the voltage is increased the nonlinearity begins to set
in which causes the peak of the forward flow to move toward lower
frequency as predicted by dilute solution theory.  As the voltage is
further increased the steric effects begin to become important with
the reversal eventually setting in.

The change in frequency of the maximum forward flow at higher voltage
as $\nu$ changes is easily understood. As $\nu$ increases, the value
of the maximum value of the capacitance decreases, shifting the peak
response to higher frequencies.  For a given $\nu$, as voltage
continues to increase, the continual decrease in the double layer
capacitance can be seen in the shift of the maximum of the positive
flow to higher frequency at high voltage. When there are no steric
effects (or Stern layer) the capacitance diverges, which can be seen
in Figure \ref{fig:steric_nu}(d).  The inclusion of the constant Stern
layer capacitance causes this decrease to limit at the constant Stern
layer capacitance \cite{olesen2006}.

\begin{figure}
\centering
(a)\includegraphics[width=2.8in]{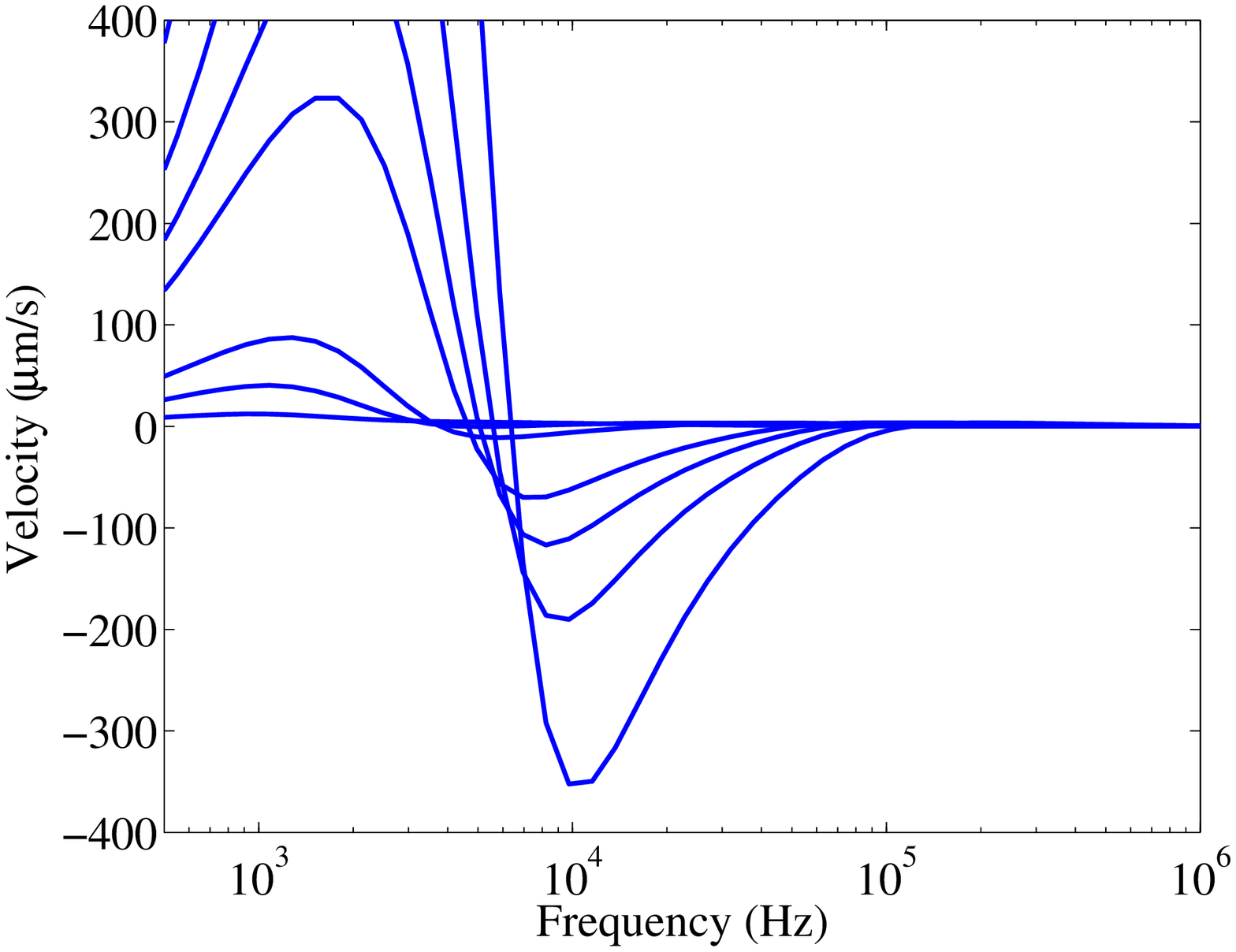}
(b)\includegraphics[width=2.8in]{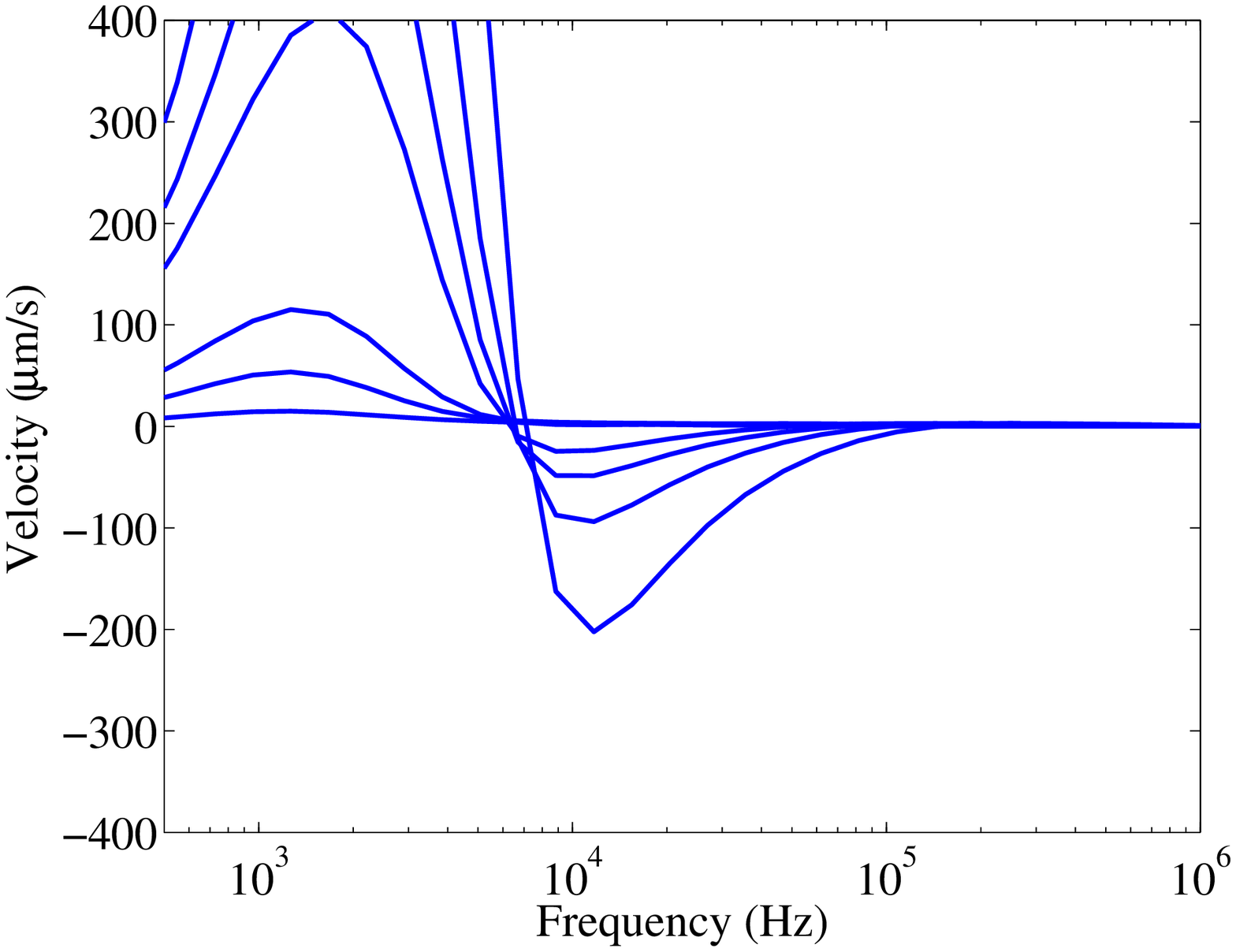}

     \caption{Flow as a function of frequency at different voltages. The voltages are
       $V = 0.5,1,1.5,2.9,3.6,4.5,$ and $6$ Volts RMS. The voltages are
       selected to match Figure 6 in Studer {\em et al.} \cite{studer2004}.
       The data is presented in dimensional form in order to 
       compare the frequency response.  
       The flow velocity is corrected for the hydraulic resistance of the
       experimental flow loop. In (a), using the Bikerman model, we 
       used $a=4.4$ nm. 
       In (b), using the Carnahan-Starling model, we 
       used $a=2.2$ nm. The general shape of both frequency response
       curves have many of the experimentally observed features.  
     }
     \label{fig:studer}
\end{figure}

Qualitatively, our model can predict a flow 
response that looks very similar to experimental
results, however the quantitative ability of our model is still problematic. 
In Fig.  \ref{fig:studer}(a) we show the model predicting
the experimental data of
Studer {\it et al.} \cite{studer2004}. In  Fig. \ref{fig:studer} we treat
the parameter  $\nu$ as a fitting parameter to approximate the
data of Figure 6 in Studer's experimental work \cite{studer2004}. We
find a value of $\nu=0.01$
provides a reasonable fit to their experimental data.
However, given a concentration of 0.1 mM KCl,  this value 
corresponds to   $a=4.4$ nm, about
an order of magnitude larger than we would expect based on the hydrated
radius of the ion. 
The fact their experiments \cite{studer2004}
only predicts reverse flow at high voltage (their Fig. 7) 
is consistent with the low
frequency regime of their work being unaccessible due to
electrochemical reactions and degradation of the electrode (see their Fig. 4).

We considered whether the Carnahan-Starling (CS) form of the excess
chemical potential would provide a more reasonable value of $a$.  In
order to include the CS equation in the model we numerically solve for
the relationship of $\Psi_D(q)$ to replace Eq. \ref{eq:q_full} in our
model. We follow the work of Bieshuvel and Soestbergen
\cite{biesheuvel2007} to obtain this relationship numerically, as
described in the Appendix.  The result of using the CS equation in our
ACEO model is shown in Fig. \ref{fig:studer} (b), and it is
qualitatively similar to our prior results with the Bikerman
model. The CS steric model has somewhat different quantitative
behavior but still requires an ion size of approximately $a=2.2$ nm to
fit the experimental data.  While the magnitude of the flow reversal
changes with the CS equation, the use of a more accurate model for the
excess chemical potential still require a relatively large ion size.

\section{Discussion}
In order to fit the experimental data, the ACEO models modified for
steric effects imply an ion length scale, $a$, almost an order of
magnitude larger than we would expect.  A factor of 10 in ion size
results in a factor of 1000 in $\nu$. This difference means that while
the response in Fig. \ref{fig:steric_nu}(a) is closer to what is
observed, it is Fig. \ref{fig:steric_nu} (c) that is predicted based
on a reasonable hydrated radius for the ions.  The current model
cannot yet predict the experimental data from first principles.  It is
important to note that the large value of $\nu$ is needed not to
predict flow reversal at all, but to predict flow reversal at the
frequencies observed in experiments.

As discussed in section ~\ref{sec:steric}, the mean-field models of
steric effects which we consider are unlikely to accurately capture
the dynamical formation of a condensed layer on the surface. One
reason is the difficulty in describing the jamming transition, as ions
and solvent molecules are squashed onto the surface by large,
time-dependent normal electric fields. Under such extreme conditions,
we should not expect more than qualitative trends from such
models. Another possibly crucial feature missing in these models is
the interaction with the surface, which is assumed to be a
mathematically flat, homogeneously charged, perfectly rigid wall, as
sketched in Fig. 2. Since the condensed layer of (presumably solvated)
ions is at the scale of only a few molecular lengths, there must be a
strong coupling with nanoscale roughness on the surface. This may have
the effect of increasing the effective ion size, as we have inferred,
since what matters is not the ion density in the solution, but the
density very close to the surface, which is reduced by atomic
asperities. The viscosity may also diverge near the surface as ions
become jammed among surface inhomogeneities~\cite{large}. It would be
interesting to do experiments with atomically flat electrodes
(e.g. carbon graphene) or ones with controlled nanoroughness to test
this hypothesis.

It is also possible that changes in permittivity, $\varepsilon$,
within the double layer may also effect the frequency response.
Electrochemists infer for the Stern layer that $\varepsilon$ may be
$1/10$ of $\varepsilon$ in the bulk ~\cite{bockris_book}.  If it were
true that $\varepsilon$ is reduced by $1/10$ in the condensed layer,
the overall double layer capacitance would be decreased.  Such change
would cause the peaks in the frequency response to shift to higher
frequency for the same value of $\nu$.  If we make the assumption that
$\varepsilon$ is reduced by a factor of ten in the {\em entire} double
layer, then an ion size of approximately 1 nm is needed to predict
Studer {\em et al.}'s \cite{studer2004} data with the CS equation.  If
$\varepsilon$ is reduced only in the condensed layer, the true effect
on the frequency response would be less than predicted with the simple
assumption. Therefore, it seems likely that accounting for changes in
$\varepsilon$ would reduce the needed ion size to correlate the data,
however at most this would reduce the ion size to 1 nm, still much
larger than expected.

A further complication is that the mean-field approximation breaks
down when ion spacings approach the Bjerrum length, $l_B =
(ze)^2/4\pi\varepsilon kT$, which is 7~ \AA ~ for bulk water and
monovalent ions ($z=1$).  Again, if it were true that $\varepsilon$ is
reduced by 1/10 in the condensed double layer, then $l_B=7$ nm.
Correlation effects on electro-osmotic flow (which to our knowledge
have never been studied) could be very significant at large voltages,
even in dilute bulk solutions~\cite{large}.

While much work remains, we have highlighted an interesting and
practically important application where dilute solution theory clearly
breaks down.  We have demonstrated that accounting for steric effects
in double-layer models can have a dramatic impact on the predicted
flow in ACEO pumps.  Because of the large value of $a$ needed to fit
the experimental data, we cannot claim that our model definitively
explains high-frequency flow reversal in ACEO pumps, but it is the
only plausible explanation to date.  The prediction is also quite
robust: we have found that flow reversal can be predicted by any model
which accounts steric effects by the generic feature that the double
layer capacitance decreases with applied voltage.  

Physically, it seems that this generic effect of decreasing
capacitance must be true for any model of the double layer considering
finite sized ions~\cite{kilic2007a}.  As the voltage increases, more
ions are packed in the double layer. Once the concentration of ions
becomes so high that the packing density is reached, the ions would
have to pile up forming a condensed layer whose thickness would
increase with voltage. As the double layer thickness grows, the
capacitance must decrease.  We believe this general effect is true
regardless of the details of the model, and we have shown that it has
major implications for ACEO pumping and, by extension, all nonlinear
ICEO flows. Steric effects are clearly among the new effects which
must be considered in developing more accurate theories of ICEO flow
at large applied voltages~\cite{large}.

\section*{Acknowledgments}
This work was supported in part by the National Science Foundation
under contract DMS-0707641 and by the U.S. Army through the Institute for
Soldier Nanotechnologies, under Contract DAAD-19-02-0002 with the
U.S. Army Research Office. MZB also acknowledges support from ESPCI
through the Joliot Chair.

\section*{ Appendix: Modified double-layer models }

In dilute solution theory, the chemical potential of a point-like ion
$i$ takes the ideal form,
\begin{equation}
  \mu_i^{ideal} = kT\log c_i + z_i e \phi \label{eq:muideal}
\end{equation}
where $c_i$ is the mean concentration and $\phi$ is the
self-consistent mean electrostatic potential, both continuous
functions in a large dilute system. In concentrated solutions, the
chemical potential is modified by a variety of statistical effects,
such as electrostatic correlations (beyond the mean-field
approximation) and interactions among discrete, finite-sized ions and
solvent molecules. In order to isolate such effects, it is customary
to decompose the chemical potential into ideal and excess
contributions, $\mu_i = \mu_i^{ideal} + \mu_i^{ex}$. Various models
for $\mu_i^{ex}$ are reviewed in
Refs.~\cite{kilic2007a,kilic2007b,biesheuvel2007}.

In the asymptotic limit of thin double layers, the chemical potential
is constant in the normal direction under rather general
conditions~\cite{chu2007}. In many cases, the algebraic conditions
$\{\mu_i=\mbox{constant}\}$ then suffice to determine the
concentration profiles $\{c_i\}$ in the diffuse layer in terms of the
mean potential distribution $\phi$. For example, in an ideal dilute
solution (\ref{eq:muideal}), the condition $\mu_i=$constant yields the
Boltzmann distribution Eq. (\ref{eq:cminus}). 

In the general case, solving the equations $\{\mu_i=\mbox{constant}\}$
and substituting the charge density $\rho(\phi)$ in Poisson's equation
leads to a modified Poisson-Boltzmann (MPB) equation for the mean
potential,
\begin{equation}
-\varepsilon \phi^{\prime\prime} = \rho(\phi) = \sum_i z_i e c_i(\phi)
\end{equation}
which can be integrated to obtain
the differential capacitance of the diffuse layer,
\begin{equation}
  C_D(\Psi_D) = \rho(\Psi_D)
  \sqrt{\frac{2\varepsilon}
{ \int_0^{\Psi_D} \rho(\phi)d\phi} }
\end{equation}
in terms of $\rho(\phi)$.  This procedure can be carried out
analytically for the Gouy-Chapman (PB) and Bikerman models to obtain
Eqs. (\ref{eq:cdpb}) and (\ref{eq:cdnu}), respectively, or numerically
for more complicated models.

There are a number of models for the excess chemical potential due to
steric effects of finite (typically hydrated) ion sizes. Here, we list
a few cases, following Biesheuvel and van
Soestbergen~\cite{biesheuvel2007}. Bikerman's model corresponds to the
continuum limit of a lattice gas for cubic ions of size $a$. In terms
of the volume fraction, $\Phi = \sum_i a^3 c_i$, the excess chemical
potential is simply~\cite{kilic2007b,biesheuvel2007},
\begin{equation}
\mu_i^{ex} = - kT \ln(1-\Phi)
\end{equation}
which comes from the entropy solvent molecules (empty lattice sites).
A more accurate expression for a liquid is the Carnahan-Starling
(CS) equation~\cite{carnahan1969},
\begin{equation}
\mu_i^{ex}=\frac{\Phi(8-9\Phi+3\Phi^2)}{(1-\Phi)^3}
\end{equation}
which has been closely validated by simulations of the hard sphere
liquid~\cite{lue1999}. There are also analytical extensions to
mixtures of hard spheres of different sizes, which have been applied
to volume effects in multicomponent
electrolytes~\cite{biesheuvel2007}.

\end{document}